\documentclass[prb,twocolumn,showpacs,amsmath,amssymb]{revtex4}
\usepackage{graphicx}
\usepackage{dcolumn}
\usepackage{bm}
\begin{document}
%JT: Modified title
\title{The nature
%JT added these two words
and boundary
%EJT
of the floating phase in a dissipative
Josephson junction array }
\author{Sumanta Tewari} \affiliation{Institute for Physical Science
and Technology and Department of Physics,
University of Maryland, College Park, MD 20742}
\author {John Toner} \affiliation{Department of Physics and
Institute of Theoretical Science, University of Oregon, Eugene, OR
97403}
\author{Sudip Chakravarty}
\affiliation{Department of Physics, University of California, Los
Angeles, Los Angeles, CA 90095-1547}
\date{\today}

\begin{abstract}
%%JT replaced this:
%The nature of
%the correlations in the floating phase of a
%dissipative Josephson junction array is clarified by detailed
%calculations of the important correlation functions.
% For the order
%parameters on the grains, it is shown that there are finite long
%range correlations in time, but only short range correlations in
%space. The dissipative transition between the floating phase and a
%global superconductor is further studied by  perturbative
%renormalization group in general dimensions.
% The transition is found to be
%with this:
We study the nature of correlations within, and   the transition
into, the floating phase of dissipative Josephson junction arrays.
Order parameter correlations in this phase are long-ranged in
time, but only short-ranged in space. A perturbative RG analysis
shows that, in {\it arbitrary} spatial dimension, the transition
is
%EJT
 controlled by a continuous locus of
 critical fixed points
 determined entirely by the
\textit{local} topology of the lattice. This may be the most
natural example of a line of critical points  existing in
arbitrary dimensions.
\end{abstract}


\maketitle
\section{Introduction}
Systems where phases and phase transitions specific to a
particular dimension are hidden in a higher dimensional manifold
have recently come under renewed focus.
\cite{Toner1,Emery,Troyer,Sumanta1} Of special importance is the
case where the hidden phases and transitions are governed by
spatially local, or zero-dimensional, modes. \cite{Sumanta1,
Sumanta2, CIKZ, Varma} A higher dimensional manifold with such
local systems embedded in it may retain some of the properties
that are effectively zero-dimensional.

%JT I insrted a paragraph break here

\begin{figure}[htb]
\centerline{\includegraphics[scale=0.4]{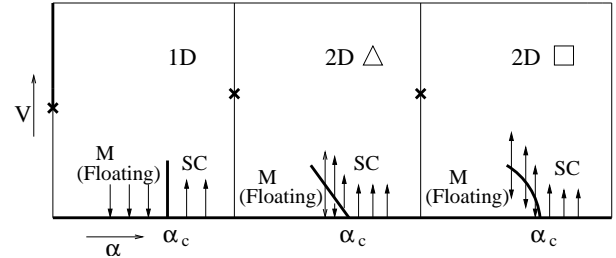}} \caption{Phase
diagram and perturbative RG flows for 1D, 2D triangular lattice
(3D face-centered-cubic), and 2D square lattice (3D simple cubic).
M and SC indicate metal and superconductor, respectively. The
thick lines correspond to line of fixed points. Crosses indicate
special points on the $\alpha=0$ axis denoting ($D+1$)-dimensional
$XY$ transition.}\label{fig:flow}
\end{figure}

It has been shown recently that this is indeed the case in a
dissipative Josephson junction array in any dimension.
\cite{Sumanta1, Sumanta2} For a system of Josephson-coupled
superconducting grains in a metallic substrate, the dissipation is
modelled by shunting resistors across the Josephson junctions.
\cite{Caldeira}
%
%
%
%JT I propose replacing this:
%
%
%In the parameter space
%comprising the coupling constant for dissipation, $\alpha$, and
%the nearest-neighbor Josephson coupling $V$, for $\alpha$ less
%than a critical value $\alpha_c$ (less dissipation), $V$ is
%irrelevant, so the phases are quantum disordered and  the system
%is metallic \cite{CIKZ}. However, for $\alpha > \alpha_c$, $V$
%grows to strong coupling and the system orders into a
%superconductor. The critical $\alpha_c$ is a quantum critical
%point.
%
%
%
%with this:
%
%
%
The phase diagram of this model in the dissipation
($\alpha$)-nearest-neighbor Josephson coupling ($V$) plane shows a
``floating'' phase at small $\alpha$ and $V$ in which $V$ is
``irrelevant'' in the renormalization group sense, so the phases
of the superconducting order parameters on different grains are
quantum disordered and  the system is metallic. \cite{CIKZ} For
larger $\alpha$ {\it or} $V$, $V$ grows to strong coupling and the
system orders into a superconductor. This phase diagram is shown
in Fig.~\ref{fig:flow}.

In addition, it has been shown recently that all longer-range
Josephson couplings, that may couple the spatially separated
junctions in the array,  are irrelevant in the metallic phase as
long as $V$ is irrelevant. \cite{Sumanta1} However, they all
become relevant simultaneously with $V$ at the same value of
dissipation, $\alpha_c$, so that there are no further transitions
within the ordered phase; the ordered phase is a global
superconductor. It has also been indicated that in the disordered
metallic phase, the superconducting grains have long range in
time, but only short range in space, correlations at zero
temperature. \cite{Sumanta1} The metallic phase has been called
`floating', the long-time-correlated zero-dimensional systems on
the grains `float' over each other.

   In the present  paper we build on our earlier work \cite{Sumanta1} by explicitly
   calculating the important correlation functions in the
   floating phase. By carefully including  all relevant and
   dangerously irrelevant variables, we show that the long-time correlation
   functions of the order parameters on individual grains are
   indeed finite,
   whereas the spatially long-range correlation
   functions are exactly zero at zero temperature.
   At nonzero temperatures, the spatial correlation functions are
   exponentially decaying with a finite
   correlation length. Thus, each grain constitutes a dynamic
   local system on its own, and they are only
   short-range-correlated in space. It is then clear that the
   dissipation-driven
   quantum phase transition (QPT) between the  metallic floating phase and
   the global superconductor is governed by the same ($0+1$)-dimensional
   physics
   as in the `quantum to classical' phase transition
   studied in the context of a single quantum
   particle in a double well \cite{Chakravarty1, Bray} or
   cosine-periodic \cite{Schmid,Zwerger} potential.
   Above $\alpha=\alpha_c$, the quantum fluctuations of
   the individual Josephson phases are effectively quenched by dissipation.
   Hence, in the absence of any quantum fluctuations,
    any infinitesimal $V$
   orders the extended system into a global superconductor.

   We further study this phase transition by computing the
   relevant renormalization group (RG) equations up to third order
   in $V$. While linear
   order recursion relations for $V$ and
   $\alpha$  only
   show the existence of the critical $\alpha_c$ to separate the
   metal from the superconductor,\cite{CIKZ, Fisher, Sumanta1} the higher order corrections
   reveal a continuous locus of critical points in the
   ($V-\alpha$)
   plane. This locus (see Fig.~\ref{fig:flow}) separates
   the metal ($V$ irrelevant) from the
   superconductor ($V$ relevant),
   determining a phase boundary on the
   ($V-\alpha$) plane that approaches the $V$-axis for
   the spatial dimension $D>1$. The bending of the phase boundary
   toward the $V$-axis is also what is physically expected and experimentally
   observed. \cite{Takahide} The functional
   form of the locus is determined entirely by the local topology of the
   lattice and does not depend on the global dimension.
   If the number of sides of the minimum closed loop on the lattice is three, for
   example in the
   $2D$ triangular or $3D$ face-centered-cubic  lattice, the locus
   is a straight line making an angle with the
   $\alpha=0$ line close to $V=0$. For the number of sides in the
   minimum closed loop greater than three, for example in $2D$
   square or $3D$ simple cubic  lattice, the locus is a
   parabola close to $V=0$. Since the forms of these loci do not
   depend on the global dimension in any way, they exist in any
   $D$; this is a demonstration of a locus of critical points in any
   dimension; there is no upper-critical-dimension for this
   problem. These results point to the intriguing possibility
   that the global transition in an extended system is ultimately
   driven by local physics at the level of a single
   junction.
    For $D=1$, the line of critical points remains vertical,
   parallel to the $\alpha=0$ line. This is analogous to
   the corresponding result in the zero-dimensional problem.
   \cite{Zwerger}  Apart from
  intrinsic theoretical interest, these results also help unravel
   the nature of the floating phase and experimental observations
   thereof. Local quantum critical points and their implications  have also
been widely discussed in the context of the cuprate
high-temperature superconductors \cite{Varma1} and heavy-fermion
materials. \cite{Si}

In section~\ref{sec:model}, we will give the preliminaries by
introducing the model and sketching the first order recursion
relation for $V$. This will enable us to quickly identify the
metallic floating phase as the one with $\alpha<\alpha_c$, and the
superconductor with $\alpha>\alpha_c$. In section~\ref{sec:corr},
we will discuss the correlation functions in the floating
phase. Some of the details
are relegated to appendix~\ref{app:correlation}. In
section~\ref{sec:secondorder}, we will compute the corrections to
the linear order recursion relation for $V$ and derive the forms
of the  loci of critical points for the triangular and square
lattices in $D=2$. Together with appendix~\ref{app:RG}, which
addresses the contribution to the flows from third order cumulant
expansion of the partition function, this completes the analysis
of the perturbative renormalization group up to third order in $V$
for the problem. The results are summarized in
Fig.~\ref{fig:flow}, which also determines the phase boundary. In
section~\ref{sec:experiments}, we will discuss possible
experimental consequences of our results. Finally, the paper is
summarized and concluded in section~\ref{sec:conclusion}.

\section{The model }
\label{sec:model}
 We start with the following action describing a Josephson
junction array coupled to dissipation in any dimension,

\begin{eqnarray}
{\cal{S}}/\hbar&=&\int_{0}^{\beta}\Bigl[\frac{C}{2}\sum_i(\frac{\partial
\theta_i}{\partial \tau})^{2} + V\sum_{\langle
i,j\rangle}[1-\cos\Delta\theta_{ij}(\tau)]\Bigr]d\tau\nonumber\\
&+& \frac{\alpha}{4 \pi}\sum_{\langle
i,j\rangle}\sum_{n}|\omega_{n}||\Delta
\tilde{\theta}_{ij}(\omega_{n})|^{2}\label{action1}
\end{eqnarray}
Here the sum $\langle i,j \rangle$ is over nearest-neighbor pairs,
and we will refer to $\Delta\theta_{ij}=(\theta_i -\theta_j)$ as
the phase difference on the bond $\langle i,j \rangle$.
  $\Delta
\tilde{\theta}_{ij}(\omega_{n})$ is a Fourier component of
$\Delta\theta_{ij}(\tau)$, and $\omega_n=\pm\ 2\pi n/\beta$, with
$n$ an integer and $\beta$ the inverse-temperature. $C$, the
capacitance of the superconducting grains, is assumed to be small,
and so $E_0\sim 1/C$ constitutes the largest energy scale in the
problem. The dimensionless variable $\alpha=R_Q/R$, where $R$ is
the  shunt resistance and $R_Q=h/4e^2$ is the quantum of
resistance, couples the Josephson-phases $\Delta\theta_{ij}$ to a
local dissipative heat bath, with a Ohmic dissipation.
\cite{Caldeira}

To establish the existence of a dissipation-tuned quantum phase transition (QPT), we employ a
frequency-shell RG which is perturbative in $V$. Let's divide the
field $\theta_i(\tau)$ into slow and fast components
$\theta_{is}(\tau)$ and $\theta_{if}(\tau)$, such that $
\theta_{is}(\tau)=\frac{1}{\sqrt \beta}\sum_{|\omega_n| \leq
\omega_c/b }\tilde{\theta}_i(\omega_n)e^{i\omega_n \tau}, $ and
$\theta_{if}(\tau)$ is given by a similar expression where the sum
is between $ \omega_c/b<|\omega_n| \leq\omega_c$. Here, $\omega_c
\sim E_0$ is the high-frequency cut-off of the frequency
integrals, and $b>1$ is a frequency rescaling factor.

We can now write the partition function $Z$ as a functional
integral,
\begin{eqnarray}
Z&=&N\int\prod_i{\cal D}\theta_i(\tau)e^{-{\cal
S}/\hbar}\nonumber\\
&=&N^{\prime}\int\prod_i{\cal D}\theta_{is}(\tau)e^{[-{\cal
S}_{0}^s/\hbar + \ln\langle e^{-{\cal S}^{\prime}/\hbar}\rangle_{0f}]}
\label{partition}
\end{eqnarray}
Here $N$ and $N^{\prime}$ are normalization constants, $S_{0}^{s}$
is the slow-frequency component of the quadratic part of the
action containing the first and the third terms in
Eq.~(\ref{action1}), $S^{\prime}$ contains the second term, and
$\langle...\rangle_{0f}$ denotes average with respect to the fast component of
the quadratic part. After computing the averages, we rescale
$\tau$, $\tau'=\tau/b$, to restore the original frequency cut-off,
and then redefine the coupling-constants to complete the
renormalization.  The dissipation term is dimensionless, and so it
is held fixed in RG. The first term in  Eq.~(\ref{action1}) then
has $\tau$-dimension $-1$. Since it is not renormalized by any of
the other terms, the coupling-constant $C$ renormalizes to zero by
power counting. As we will show in the next section, we can put
this term equal to zero for the purposes of the RG, since
inclusion of it does not change any of the results. For
computations of some correlation functions in the floating phase,
however, $C$ is dangerously irrelevant; some of these functions
will depend singularly on $C$.

 The average in Eq.~(\ref{partition}) can be easily performed in
the leading order in $V$,
\begin{eqnarray}
&&\langle\exp [V\sum_{\langle i, j \rangle}
\int_0^{\beta}d\tau\cos\Delta\theta_{ij}(\tau)]\rangle_{0f}\nonumber\\&=&
\exp[\frac{1}{2}V\sum_{<ij>} \int_0^{\beta}d\tau
\sum_{\epsilon=\pm 1}e^{i\epsilon\Delta\theta_{ij}^s(\tau)}
\langle e^{i\epsilon\Delta\theta_{ij}^f}(\tau)\rangle_{0f}]
\nonumber\\&=&\exp[V\sum_{\langle i, j \rangle}
\int_0^{\beta}d\tau
\cos\Delta\theta_{ij}^s(\tau)e^{-\frac{1}{2}\langle
\Delta\theta_{ij}^f(\tau)^2\rangle_{0f}}],
\end{eqnarray}
where,
\begin{eqnarray}
\langle\Delta\theta_{ij}^f(\tau)^2\rangle_{0f}&=&\frac{1}{z_0}\frac{1}{\beta}\sum_{\omega_c/b
<|\omega_n|\leq\omega_c}\frac{1}{
\frac{\alpha}{2\pi}|\omega_n|}\nonumber\\
&=&\frac{2\ln b}{z_0\alpha} \label{average1}
\end{eqnarray}
Here $z_0 = z/2$, where $z$ is the coordination number of the
lattice. In deriving Eq.~\ref{average1} we have taken $C$ to be
zero, which, as mentioned before, is allowed for the present
purpose.
  After rescaling the time, $\tau'=\tau/b$, and taking $b=e^l$, where
  $l>0$ is infinitesimal, we end up with a
  linear order RG flow equation for
$V$,
\begin{equation}
\frac{dV}{dl}=(1-\frac{1}{z_0\alpha})V \label{Vrg}
\end{equation}
It is clear from this equation that $\alpha=\alpha_c=1/z_0$ is a
critical fixed point for the flow of $V$; for $\alpha<\alpha_c$,
$V$ scales to zero, while for $\alpha > \alpha_c$, $V$ grows. In
the phase where $V$ scales to zero, the junction-phases are
quantum-disordered and the system is metallic by construction due
to the existence of the shunt resistors. When $V$ flows to higher
values, the system is phase-ordered and superconducting.

Now we show that in the metallic phase, all longer-ranged
Josephson couplings beyond the nearest-neighbor coupling $V$,
which can couple spatially separated $\Delta\theta_{ij}$'s on the
array if included in the starting action, are irrelevant. These
couplings are described by the general term,
  \begin{equation}
{\cal{S}}_J/\hbar=-\sum_{[s_i]}\sum_{i}\int_0^{\beta}d\tau
J([s_i])\cos[\sum_j s_j\theta ({\vec r}_i + {\vec \delta}_j,
\tau)] \label{action2}
  \end{equation}
Here, $\theta ({\vec r}_i, \tau) $'s are the order parameter
phases in the  superconducting grains, the ${\vec \delta}_i$'s are
arbitrary vectors separating the lattice points on the lattice
%(to
%allow for arbitrary ranged couplings)
and $s_i$ is an integer-valued function of the layer number $i$
satisfying $\sum_i s_i = 0$. The last constraint ensures the
absence of an external field, which implies ``rotation invariance"
under adding the same constant to all of the $\theta$'s. Note that
the special case $s_0 =+1 , {\vec \delta}_0 = {\vec 0}, s_1 =-1 ,
{\vec \delta}_1 = {\vec a_{\gamma}}$, where ${\vec a_{\gamma}}$ is
a nearest neighbor vector, is just the nearest neighbor Josephson
coupling $V$ in Eq.~(\ref{action1}).

Calculations  analogous to that leading to Eq.~(\ref{Vrg}) give,
to first order in $J$,
\begin{equation}
\frac{dJ([s_i])}{dl}=[1-\frac{\Gamma([s_i])}{\alpha}]J([s_i]),
\label{Jsirg}
\end{equation}
  with
\begin{equation}
\Gamma([s_i])\equiv \sum_{i,j }  s_i s_j U( {\vec \delta}_i -
{\vec \delta}_j). \label{Gammadef}
\end{equation}
The ``potential'' $ U( {\vec r}) \equiv \frac{1}{N}\sum_{{\vec
q}}(e^{i{\vec q} \cdot {\vec r}} - 1)/f_{\vec q}$, with $f_{\vec
q} \equiv  \sum_{{\gamma}}(1 - e^{i{\vec q} \cdot {\vec a_\gamma}}
)$, where the sum over $\gamma$ is over all nearest neighbors. It
is straightforward to show that $U({\vec r})$
  is  the ``lattice Coulomb potential''
  of a unit negative charge at the origin, with the zero of
  the potential set at ${\vec r} = {\vec 0}$. That is, $U({\vec r})$ satisfies
  the ``lattice Poisson equation":
  $ \sum_\gamma U({\vec r} - {\vec a}_\gamma) - z U ({\vec r}) = -
\delta_{{\vec r}, {\vec 0}}$.
  For a symmetrical (e.g., square, hexagonal, cubic) lattice, the
left-hand side is
  just the ``lattice Laplacian'', approaching $\nabla^2 U \times O(a^2)$ where
  ($a \equiv |{\vec a}_\gamma |$).

The quantity $\Gamma$ in (\ref{Jsirg}) and (\ref{Gammadef}) is
then just equal to the potential energy of a neutral (since
$\sum_i s_i = 0$) plasma of quantized (since the $s_i$'s are
integers) charges $s_i$ on the lattice. The most relevant
$J([s_i])$ is  clearly the one that corresponds, in this Coulomb
analogy, to the lowest interaction energy. Note that strictly
speaking $\Gamma$ corresponds to {\it twice} this energy, because
the sum in (\ref{Gammadef}) double counts. Apart from the trivial
configuration in which all the $s_i = 0$ , the lowest energy
configuration is clearly the one in which there are two equal and
opposite unit magnitude charges on nearest neighbor sites: i.e.,
$s_0 =+1  , {\vec \delta}_0 = {\vec 0}, s_1 =-1 , {\vec \delta}_1
= {\vec a_\gamma}$. As we discussed earlier, this corresponds to
the nearest-neighbor Josephson coupling in equation
(\ref{action1}). Thus, it is established that that coupling is,
indeed, the most relevant, as we asserted earlier. Furthermore,
using simple symmetry arguments, we can show that for a symmetric
lattice (e.g., square, hexagonal, cubic), where all
nearest-neighbor sites are equivalent, $U({\vec a}_\gamma ) = -
\frac{1}{2 z_0}$, which recovers the recursion relation
(\ref{Vrg}) for $V$. As a result, all other couplings are {\it
irrelevant} for $\alpha \leq \alpha_c = 1/z_0$; hence, they affect
neither the floating phase nor the transition between it and the
($D + 1$)-dimensionally coupled phase.

Thus, the metallic phase is in a sense spatially decoupled. This
decoupling of space, but finite correlation in time, can be more
clearly seen by the correlation functions in the floating phase
discussed in the next section.

\section{Correlation functions in the floating phase}
\label{sec:corr} In this section, we first show that the
variable $C$ is dangerously irrelevant in $D\leq 2$ for some
correlation functions in the floating phase, but not for the ones
required for the RG. Including all relevant and dangerously
irrelevant variables, we will then compute various long-time
correlation functions (Eqs.~\ref{corr1}, \ref{corr2}, \ref{corr3},
 \ref{corr4}) and show that they are  finite. Finally, the
spatially long-range correlation functions (Eqs.~\ref{3Dopcorr},
\ref{2Dopcorr}, \ref{1Dopcorr}) will be shown to be exponentially
decaying with finite correlation lengths at a non-zero
temperature and exactly zero in the limit of zero
temperature. Thus, the intrinsically  local and dynamic character
of the floating phase will be established, indicating that the
   dissipation-driven
    QPT between this phase and
   the global superconductor is governed by the same ($0+1$)-dimensional
   physics as in the corresponding zero-dimensional
   transition. \cite{Chakravarty1, Bray, Schmid,Zwerger}

From the gaussian part of the action in Eq.~\ref{action1}, we find that
\begin{equation}
\langle|\theta({\bf k} \omega_n)|^2\rangle=\frac{1}{C\omega_n^2 + (\alpha/2\pi)f({\bf
k})|\omega_n|}. \label{propagator}
\end{equation}
Here $f({\bf k})=\sum_{\mathbf {\delta}}|(e^{i{\bf k}.{\mathbf
{\delta}}}-1)|^2$, where ${\mathbf{\delta}}$'s constitute the set
of nearest neighbor vectors. Since $f({\bf k})\sim z_0 k^2$ for
small ${\bf k}$, it is easy to see that in dimension $D\leq 2$,
wavenumber integration of the right-hand-side (RHS) of
Eq.~\ref{propagator} diverges in the infrared for $C\rightarrow
0$. This divergence forces us to include a non-zero $C$ for the
correct evaluation of some of the correlation functions in the
floating phase, even though $C$ is an irrelevant variable in the
RG sense. The recursion relations, however, remain unaffected as
we show below.

The only correlation functions we need for the purposes of the RG
involve the variables defined on a bond,
$\Delta\theta_{ij}(\tau)$, the simplest of which is given in
Eq.~\ref{average1}, where we have taken $C$ to be zero. In
calculations higher order in $V$, we will need higher order
correlators $\sim
\langle\Delta\theta_{ij}(\tau)\Delta\theta_{kl}(\tau)\rangle_{0f}$, still
among the bond-variables. For such functions, it is easy to see
that including a non-zero $C$ on the RHS of Eq.~\ref{propagator}
only amounts to introducing a high-frequency cut-off $E_0\sim 1/C$
in the frequency integrals. In the case of
$\langle \Delta\theta_{ij}(\tau)^2\rangle$, summing over $j$ over the
nearest-neighbor vectors assuming symmetry among the nearest
neighbors, dividing by $z_0$, and using Eq.~\ref{propagator}, we
get in $D=1$,
\begin{eqnarray}
\langle\Delta\theta_{ij}(\tau)^2\rangle&=&\frac{1}{z_0}\int_{0}^{\Lambda}\frac{dk}{2\pi}
\sum_{\frac{1}{\beta}\leq|\omega_n|}f({\bf k})\langle |\theta({\bf
k}\omega_n)|^2\rangle\nonumber\\&=&\frac{1}{z_0}\int_{0}^{\Lambda}\frac{dk}{2\pi}
\sum_{\frac{1}{\beta}\leq|\omega_n|}\frac{f({\bf k})}{C\omega_n^2+
(\alpha/2\pi) f({\bf k})|\omega_n|},\nonumber
\end{eqnarray}
 where $\Lambda$ is a momentum cut-off. It is easy to see that
 since $f({\bf k})$ appears both in the numerator and the
 denominator, the infrared divergence in the wavenumber integration is
 eliminated. Since in the limit of small $k$ the integral is zero,
  $f({\bf k})$ is order one in the integral,  hence $C\omega_n^2$
  can be neglected with respect to $(\alpha/2\pi)f({\bf
  k})|\omega_n|$ for small $\omega_n$. An explicit evaluation of the integral
  shows the same logarithmically divergent behavior as in
  Eq.~\ref{average1},
  $\langle \Delta\theta_{ij}(\tau)^2\rangle=\frac{2}{z_0\alpha}\ln(\beta E_0)$,
  where $E_0=K/C$ and $\ln K=\int_0^{\Lambda}\frac{dk}{2\pi} \ln (\alpha f({\bf
  k}))$. Hence, for the purposes of the RG, taking $C=0$ in the expression for $<|\theta({\bf
  k},\omega_n)|^2>$ and taking a compensatory high-frequency cut-off $\sim
  1/C$ are indeed justified.

  From  $\langle\Delta\theta_{ij}(\tau)^2\rangle$, we can
  calculate
\begin{equation}
  \langle \exp(iq\Delta\theta_{ij}(\tau))\rangle=1/(\beta
  E_0)^{\frac{q'^2}{z_0\alpha}}.
  \label{corr0}
  \end{equation}
   Here, $q'^2\equiv^{\text min}_{n\in Z}(q-n)^2 $,
   and $q$ is a real number. Note that for $q=1$, the simplest
   such function that one can construct, the RHS of
   Eq.~\ref{corr0} does not vanish at zero temperature since the
   corresponding $q'$ vanishes. This is analogous to a single
   paramagnetic spin developing a non-zero expectation value in
   the presence of an applied external field. In the present case,
   the Josephson coupling $V$ acts as the field inducing a
   non-zero expectation value of $e^{i
\Delta \theta_{ij}}$, even in the complete absence of all other
couplings. This expectation value  does not constitute an order
parameter since the broken symmetry is  not spontaneous, but
merely induced by the applied field. It forces us, however, to
introduce the parameter $q$ to get correlation functions which do
decay algebraically \cite{Sumanta1} as $T\rightarrow 0$. Note that
$q'^2$ is non-zero as long as $q$ is not an integer.

The result for $q'$ is analogous to a well-known result in surface
roughening, \cite{rough, srough} which we shall sketch here for
the sake of completeness. The result, $ \langle
\Delta\theta_{ij}(\tau)^2\rangle=2\ln (\beta E_0)/z_0\alpha$,
correctly gives the low-temperature fluctuations of $\Delta
\theta_{ij}$ throughout the floating phase, up to an unimportant
additive constant. Since this is valid irrespective of the actual
bare value of $V$, we can apply it even in the limit of large $V$.
But since in this limit $\Delta\theta_{ij}$ is quantized in
integral multiples of $2\pi$, $\Delta\theta_{ij}=2\pi n$ with $n$
an integer, we can write for the function in Eq.~\ref{corr0},
\begin{eqnarray}
&&\langle\exp[iq\Delta\theta_{ij}(\tau)]\rangle
=\sum_{n=-\infty}^{\infty}P(n)\exp(2\pi i q n)\nonumber\\&=&
\sum_{n=-\infty}^{\infty}\Big(\frac{\pi z_0\alpha}{\ln (\beta
E_0)}\Big)^{\frac{1}{2}}\exp\left[2\pi i q n - \frac{n^2\pi^2z_0\alpha}{
\ln(\beta E_0)}\right]. \label{gauss1}
 \end{eqnarray}
Here $P(n)$ is the gaussian distribution function for $n$ with
$\langle
n\rangle=\frac{1}{2\pi}\langle\Delta\theta_{ij}(\tau)\rangle=0$,
the standard deviation
$\sigma_n=\frac{1}{2\pi}\langle\Delta\theta_{ij}(\tau)^2\rangle$,
and the sum over $n$, instead of an integral over a continuous
variable, takes care of the quantization condition on
$\Delta\theta_{ij}$. Doing this sum using the Poisson summation
formula gives
\begin{equation}
\langle \exp[iq\Delta\theta_{ij}(\tau)]\rangle=\sum_{s=-\infty,
s\in  Z}^{\infty}(\beta E_0)^{-(q-s)^2/(z_0\alpha)}.
\end{equation}
 Clearly, the dominant term in this sum as $T\rightarrow 0$ is the one with
the smallest value of $( q - s)^2$. Keeping just this term yields
Eq.~(\ref{corr0}). This shows that this correlation function
indeed goes to zero as the temperature goes to zero. The
unequal-time correlation function of the bond-variable is also
algebraic,
\begin{equation}
 \langle \exp[iq(\Delta\theta_{ij}(\tau)-\Delta\theta_{ij}(0))]\rangle
=\frac{1}{(E_0\tau )^{2 q'^2/z_0\alpha}}
 \label{corr1}.
\end{equation}

The  correlation function of the bond variables algebraically
decaying in time, Eq.~\ref{corr1},  is valid in all dimensions.
The same is not true, however, with the correlation functions of
the site variables, $\theta_i(\tau)$, where $i$ is a grain index,
since now for $D\leq 2$ the wavenumber integration of
$<|\theta({\bf k},\omega_n)|^2>$ is divergent as $C\rightarrow 0$.
For $D
> 2$, the effect coming from $C$ is subleading and the long-time
correlations remain algebraically decaying. In $D=3$, for the
spatially local but unequal in time correlator we get, ignoring
unimportant constant factors,
\begin{equation}
\langle \exp[iq(\theta_i(\tau)-\theta_i(0))]\rangle\sim \frac{1}{(E_0\tau
)^{\frac{q'^2\Lambda}{\pi\alpha z_0}}}, \label{corr2}
\end{equation}
where $\Lambda$ is the momentum cut-off. In $D=2$, the same
correlation function is given by,
\begin{equation}
\langle \exp[iq(\theta_i(\tau)-\theta_i(0))]\rangle\sim
\exp\left[-\frac{q'^2}{4\pi\alpha z_0}(\ln \frac{\tau\alpha z_0}{2\pi
C})^2\right], \label{corr3}
\end{equation}
while in $D=1$ it is,
\begin{equation}
\langle \exp[iq(\theta_i(\tau)-\theta_i(0))]\rangle \sim
\exp\left[-\frac{\surd(2)q'^2}{(\pi\alpha z_0
C)^{1/2}}(\tau)^{\frac{1}{2}}\right]. \label{corr4}
\end{equation}

Even though the long-time behavior is algebraic only for the
bond-variables, Eq.~\ref{corr1}, and for the site variables only
for $D=3$ and above, Eq.~\ref{corr2}, temporally far-separated
sites are still correlated for $D\leq 2$ according to
Eqs.~(\ref{corr3}, \ref{corr4}). However, long-range spatial
correlation functions of the grains,
$\langle \exp[iq(\theta_i(\tau)-\theta_{i+r}(\tau))]\rangle$, are identically
zero at zero temperature in any dimension. Spatially far-separated
grains are completely uncorrelated, each one of them constitutes a
dynamic local system on its own. In this sense, the temporal and
spatial behaviors of the system are completely decoupled at $T=0$.
It is also a manifestation of the intrinsic local nature of the
problem. At nonzero $T$, the spatial correlation functions are
exponentially decaying in all dimensions; the grains are only
short-range-correlated.

As we show in appendix~\ref{app:correlation}, the correlation
function of the phases of the order parameters on the grains
follow the simple scaling law,
\begin{equation}
\langle \theta_i(\tau)\theta_{i+r}(\tau)\rangle =\frac{2\pi r^{2-D}}{\alpha
z_0}g(r/\xi), \label{scaling}
\end{equation}
where, $\xi$ is the correlation length at nonzero $T$,
$\xi=(\alpha z_0/2\pi CT)^{1/2}$. As we will see below, even
though $\xi$ diverges as $T\rightarrow 0$, there is very little
correlation left between the order parameters themselves at low
temperatures, and eventually the order parameter correlation
function
%,$<\exp[iq(\theta_i(\tau)-\theta_{i+r}(\tau))]>$,
 vanishes at $T=0$. In $D=3$, we find,
\begin{eqnarray}
g(r/\xi)&=& \frac{1}{2\pi^2}\frac{\xi}{r}\exp(-r/\xi),
\hspace*{0.3 in} r>>\xi\nonumber\\&=&\frac{1}{2\pi^2}\ln(\xi/r),
\hspace*{0.5 in}r<<\xi \label{3Dscaling}
\end{eqnarray}
This gives for the correlation function of the order parameters,
\begin{eqnarray}
\langle e^{iq(\theta_i(\tau)-\theta_{i+r}(\tau))}\rangle_c
&=&T^{\frac{q'^2\Lambda}{\pi\alpha z_0}}\frac{q'^2}{\pi\alpha z_0
r}\frac{\xi}{r}e^{-\frac{r}{\xi}},\hspace*{.1
in}r>>\xi\nonumber\\\hspace*{-.1
in}&=&T^{\frac{q'^2\Lambda}{\pi\alpha
z_0}}\left[(\frac{\xi}{r})^{\frac{q'^2}{\pi\alpha z_0
r}}-1\right], \hspace*{.05 in}r<<\xi\nonumber\\ \label{3Dopcorr}
\end{eqnarray}
where the subscript $c$ denotes the connected piece of the
correlation function. It is clear that as $T\to 0$, these
functions vanish for large $r$.
Hence, there is no correlation among the spatially far separated
grains at zero temperature. In $D=2$, we get for the scaling
function $g(r/\xi)$,
\begin{eqnarray}
g(r/\xi)&=&
\frac{1}{\surd(2\pi^3)}(\frac{\xi}{r})^{\frac{3}{2}}\exp(-r/\xi),
\hspace*{0.3 in}
r>>\xi\nonumber\\&=&\frac{1}{2\pi^2}[\ln(2\xi/r)]^2, \hspace*{0.7
in}r<<\xi \label{2Dscaling}
\end{eqnarray}
The order parameter correlation function becomes,
\begin{widetext}
\begin{eqnarray}
\langle\exp[iq(\theta_i(\tau)-\theta_{i+r}(\tau))]\rangle_c&=&\exp\left[-\frac{(q'\ln\frac{\beta\alpha
z_0}{2\pi C})^2}{4\pi\alpha
z_0}\right](\frac{2}{\pi})^{1/2}\frac{q'^2}{\alpha
z_0}(\frac{\xi}{r})^{3/2}\exp(-\frac{r}{\xi}),\hspace*{.5 in}
r>>\xi\nonumber\\&=&\exp\left[-\frac{(q'\ln\frac{\beta\alpha
z_0}{2\pi C})^2}{4\pi\alpha
z_0}\right]\Big(\exp[\frac{(q'\ln(r/2\xi))^2}{\pi\alpha
z_0}]-1\Big), \hspace*{0.6 in}r<<\xi \label{2Dopcorr}
\end{eqnarray}
\end{widetext}
It is straightforward to check that this function is also zero at
$T=0$. Finally, in $D=1$, $g(r/\xi)$ is given by,
\begin{eqnarray}
g(r/\xi)&=&(1/\pi)(\xi/r)^2\exp(-r/\xi), \hspace*{.4
in}r>>\xi\nonumber\\&=&-\frac{1}{\pi}\ln(\frac{\xi}{r}) +
\frac{1}{\pi}\frac{\xi}{r}\exp(-r/\xi), \hspace*{.1in}r<<\xi
\label{1Dscaling}
\end{eqnarray}
The corresponding order parameter correlation function is,
\begin{widetext}
\begin{eqnarray}
\langle\exp[iq(\theta_i(\tau)-\theta_{i+r}(\tau))]\rangle_c&=&\exp\left[-\frac{q'^2\surd
(2\beta)}{\surd(\pi \alpha z_0 C)}\right](\frac{2q'^2}{\alpha
z_0})(\frac{\xi^2}{r})\exp(-r/\xi), \hspace*{.2 in}
r>>\xi\nonumber\\&=&(r/\xi)^{(2q'^2r)/(\alpha
z_0)}-\exp\left[-\frac{q'^2\surd (2\beta)}{\surd(\pi \alpha z_0
C)}\right],\hspace*{.4 in} r<<\xi\label{1Dopcorr}
\end{eqnarray}
\end{widetext}
%As before, in the limit $T\rightarrow 0$, which corresponds to
%$r<<\xi$ in the above function,
%Noting that the disconnected piece is $(\frac{2\pi r^2 CT}{\alpha
%z_0})^{(q'^2r)/(\alpha z_0)}$,
This function also approaches zero as $T \to 0$. Thus, we have
established that the spatially far-separated grains are
uncorrelated at $T=0$, and correlated only with exponentially
decaying correlations at a non-zero temperature.

%At non-zero temperature, the spatial correlation functions  also
%decay algebraically, spatial coupling arising from $C$, although
%the exponents of the algebraic decay are simple numbers (in fact
%they are exactly equal to the spatial dimension $D$) which do not
%depend on $\alpha$. For non-zero $T$ and in $D=3$, we find,
%\begin{equation}
%<\exp[iq(\theta_i(\tau)-\theta_{i+r})]>_c=\frac{1}{(\beta
%E_0)^{\frac{q'^2\Lambda}{\pi\alpha z_0}}}\frac{\beta
%q'^2}{6\pi^2C}\frac{1}{r^3}, \label{spacecorr3}
%\end{equation}
%which is valid for large $r$, and the subscript $c$ denotes the
%connected piece of the correlation function. In $D=2$ it is given
%by,
%\begin{equation}
%<\exp[iq(\theta_i(\tau)-\theta_{i+r})]>_c=\exp[-\frac{q'^2(\ln
%\frac{\beta}{C})^2}{4\pi\alpha z_0}]\frac{\beta
%q'^2}{4\pi^2C}\frac{1}{r^2}, \label{spacecorr2}
%\end{equation}
% while in $D=1$, the exponent for the long-range decay is one,
%\begin{equation}
%<\exp[iq(\theta_i(\tau)-\theta_{i+r})]>_c=\exp[-\frac{q'^2(\beta)^{\frac{1}{2}}}{(2\pi\alpha
%z_0C)^\frac{1}{2}}]\frac{\beta q'^2}{2\pi^2C}\frac{1}{r}.
%\label{spacecorr1}
%\end{equation}
%At $T=0$, however, these correlation functions are identically
%zero in all dimensions.

\section{RG at second order in V}
\label{sec:secondorder} To establish the existence of a locus of
critical points in the $\alpha-V$ plane, we need to compute the
higher-order corrections, if any, to the first order recursion
relations for $V$, Eq.~(\ref {Vrg}). Performing the cumulant
expansion of Eq.~(\ref{partition}) in second order in $V$, we get
\begin{eqnarray}
&\exp&\Bigr[\frac{V^2}{2}\Bigr(\langle\int_0^\beta d\tau
\int_0^\beta
d\tau'\sum_{\langle i,j \rangle}\sum_{\langle k,l \rangle}\cos\Delta\theta_{ij}(\tau)\nonumber\\
&&\cos\Delta\theta_{kl}(\tau')\rangle_{0f} - \langle\int_0^\beta
d\tau
\sum_{\langle i,j \rangle}\cos\Delta\theta_{ij}(\tau)\rangle_{0f}^2\Bigl)\Bigl]\nonumber\\
&=&\exp\Bigl[e^{-\frac{2\ln
b}{z_0\alpha}}\Bigl(\frac{V^2}{4}\int_0^\beta d\tau\int_0^\beta
d\tau'\sum_{\langle i,j \rangle}\sum_{\langle k,l \rangle}\Bigl[\cos(\Delta\theta_{ij}^s(\tau)\nonumber\\
&+&\Delta\theta_{kl}^s(\tau'))(e^{-\langle\Delta\theta_{ij}^f(\tau)
\Delta\theta_{kl}^f(\tau')\rangle_{0f}}-1)+\cos(\Delta\theta_{ij}^s(\tau)\nonumber\\
&-&\Delta\theta_{kl}^s(\tau'))(e^{\langle\Delta\theta_{ij}^f(\tau)
\Delta\theta_{kl}^f(\tau')\rangle_{0f}}-1)\Bigr]\Bigr)\Bigr]\label{secondorder}
\end{eqnarray}

To avoid the generation of spurious long-ranged behavior in the
correlation function of the fast modes,
$\langle\Delta\theta_{ij}^f(\tau)
\Delta\theta_{kl}^f(\tau')\rangle_{0f}$, we need to adopt a smooth
cut-off prescription \cite{Ma} in our frequency-shell RG. For the
unequal-time correlation function on the same bond, this can be
written as
\begin{eqnarray}
&G^{f}_{ij,ij}&((\tau-\tau'),b)\equiv
\langle\Delta\theta_{ij}^f(\tau)
\Delta\theta_{ij}^f(\tau')\rangle_{0f}\nonumber\\&=&
\frac{1}{z_0\beta}\sum_{|\omega_n|<\omega_c}\frac{\exp(i\omega_n(\tau-\tau'))}{\alpha/2\pi
|\omega_n|}f(\omega_n/(\frac{\omega_c}{b})),
\end{eqnarray}
  where $f(x)$ is a smoothing function with the properties
  $f(x)\rightarrow 0$ for $x<<1$, and $f(x)\sim 1$ for $x>>1$. The choice of
  $f(x)$ is somewhat arbitrary as long as it entails the
  properties, $G^{f}_{ij,ij}((\tau-\tau'), b)\rightarrow 0$  for
  $|\tau-\tau'|\rightarrow \infty$, and $G^{f}_{ij,ij}((\tau-\tau'), b)\rightarrow
   0$ for $b\rightarrow 1$, to
  the correlation function of the fast modes. A specific choice is given in Ref.
  \onlinecite{Zwerger}, where $G$ goes to zero exponentially with $|\tau-\tau'|$ for
  large $|\tau-\tau'|$, and $G^{f}_{ij,ij}(0,b)=\frac{2\ln b}{z_0 \alpha}$.
  This does not modify
  any of our previous calculations and we will assume that
  $G^{f}_{ij,kl}((\tau-\tau'),b)$ is short-ranged in $|\tau-\tau'|$. The $\tau$ and $\tau'$
  in the integrals of Eq.~(\ref{secondorder}) are now constrained
  to be close to each other and we can use gradient-expansion in $(\tau-\tau')$to
  simplify the terms within parenthesis. It is clear that the
  second term does not renormalize $V$.
The first term, however, upon gradient expansion  generates
$\cos(\Delta\theta^s_{ij}(\tau)+\Delta\theta^s_{kl}(\tau))$ summed
over all bonds $\langle i, j \rangle$ and $\langle k, l \rangle$.

 %In one dimension, $G^{f}_{ij,kl}(\tau-\tau')=0$ for
%$\langle i, j \rangle\neq\langle k, l \rangle$. The first term in Eq.~(\ref{secondorder}) then
%produces a term $\sim \cos(2\Delta\theta^s_{ij}(\tau))$ which is
%irrelevant close to $\alpha_c$. The second term also produces an
%irrelevant term, $\sim (\frac{\partial\Delta\theta_{ij}^s(\tau)}
%{\partial\tau})^2$.
It is clear that there is no renormalization of $V$ in second
order in one dimension. In the sum over $\langle i, j \rangle$,
and $\langle k, l \rangle$, if $\langle i, j \rangle=\langle k, l
\rangle$, one generates a term $\sim
\cos(2\Delta\theta^s_{ij}(\tau))$ which is irrelevant close to
$\alpha_c$. The second term also produces an irrelevant term,
$\sim (\frac{\partial\Delta\theta_{ij}^s(\tau)}
{\partial\tau})^2$. For $\langle i, j \rangle\neq\langle k, l
\rangle$, too, it is easy to see that there is no renormalization
of $V$, since $\cos(\Delta\theta_{ij}+\Delta\theta_{kl})$ does not
produce $\cos(\Delta\theta_{mn})$; in one dimension there is no
closed loop on the lattice to do this. Contrast this with $D=2$
triangular lattice, where because of the existence of closed loops
in the lattice $\cos(\Delta\theta_{mn})$ is produced from
$\cos(\Delta\theta_{ij}+\Delta\theta_{kl})$ when the bonds
$\langle i, j \rangle, \langle k, l \rangle, \langle m, n \rangle$
form the three sides of a minimum triangle. Similar effects exist
for any lattices in $D>1$. Coming back to $D=1$, we have checked
explicitly, as shown in the appendix~\ref{app:RG} in a more
general setting, that there is no renormalization of $V$ also at
the third order. Thus we speculate, and confirm up to third order
expansion in $V$, that in $D=1$ there is no correction at all at
any higher order to the linear order recursion relation
Eq.~\ref{Vrg}. This was conjectured in the context of the
corresponding $0$-dimensional problem in Ref.~\cite{Zwerger}, here
we have extended it to one dimension. Notice that in
Eq.~\ref{action1}, because of the existence of the $C$-term, the
one dimensional problem does not map onto the zero dimensional
problem in any obvious way.
%In fact, since in
%the computation of the correlation functions bonds $\langle i, j \rangle$ and
%$\langle k, l \rangle$ decouple in $1D$ for $\langle i, j \rangle\neq\langle k, l \rangle$ , technically the
%problem maps exactly on to the zero-dimensional single junction
%problem, for which it has been argued \cite{Zwerger} that the
%recursion equation Eq.~(\ref{Vrg}) is exact for $V$. Contrast the
%case of one dimension with any $D>1$, where spatially separated
%$\langle i, j \rangle$ and $\langle k, l \rangle$ do not decouple because of the existence of
%loops.
We will see below that for $D>1$ there are corrections to
Eq.~(\ref{Vrg}) at higher orders.

In $2D$ triangular array, and in fact in arbitrary dimensions
where the minimum closed loop is a triangle, there is a
contribution at order $V^2$ to the recursion relation for $V$.
This comes from the first term in (\ref{secondorder}), where, in
the sum over $\langle k, l \rangle$, when $\langle k, l \rangle$
is the bond adjacent to $\langle i, j \rangle$ in a minimum
triangle, the sum over $\langle i, j \rangle$ and $\langle k, l
\rangle$ produces a single sum over all bonds in the lattice. This
is illustrated by Fig.(1a). If $\langle i, j \rangle=<1,2>$, and
$\langle k, l \rangle=<2,3>$, then
$\cos(\Delta\theta^s_{ij}+\Delta\theta^s_{kl})=\cos(\Delta\theta^s_{13})$
since $\Delta\theta_{12}+\Delta\theta_{23}+\Delta\theta_{31}=0$
around a  triangle. The first term within parenthesis in
Eq.~(\ref{secondorder}) can then be written as
\begin{eqnarray}
&&e^{-\frac{2\ln b}{z_0\alpha}}\frac{V^2}{4}\int_0^\beta
d\tau\sum_{\langle i, j \rangle}\cos(\Delta\theta^s_{ij}
(\tau))\nonumber\\&\times&\int_{-\tau}^{\beta}d(\tau'-\tau)[\exp(-G^{f}_{12,23}(\tau-\tau'))-1],
\label{firstterm}
\end{eqnarray}
 where $G^{f}_{12,23}(\tau-\tau')$ is the nearest-neighbor-bond
correlation function of the fast modes. For the triangle in
Fig.(1a), using
$\Delta\theta_{12}+\Delta\theta_{23}+\Delta\theta_{31}=0$, we get
$G^{f}_{12,23}(\tau-\tau')=-(1/2)G^{f}_{12,12}(\tau-\tau')$. The
precise values of these correlation functions are somewhat
arbitrary since they depend on the choice of the smoothing
function $f(x)$. Assuming they are finite only close to
$|\tau-\tau'|=0$, we can then define $\tau_0$ so that the
$(\tau'-\tau)$ integral in Eq.~(\ref{firstterm}) is given by
$\tau_0[\exp(\frac{1}{2}G^{f}_{12,12}(0))-1]$. Using
$G^{f}_{12,12}(0)=\frac{2\ln b}{z_0 \alpha}$, and rescaling
$\tau$, we find the recursion relation for $V$ at second order in
$V$
\begin{equation}
\frac{dV}{dl}=V(1-\frac{1}{z_0\alpha}) + C_1V^2,
\end{equation}
 where $C_1$ is a positive constant $C_1=\tau_0/4z_0\alpha$.
 %The
 %value, but not the sign,
 %of $C_1$ is of course not unique, and it inherits the arbitrariness
 %of the smoothing function.
\begin{figure}[htb]
 \centerline{\includegraphics[scale=0.4]{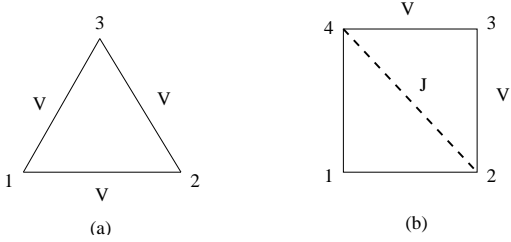}}
  \caption{The minimum closed loop generating higher order
  renormalization of
$V$ for (a) 2D triangular lattice, and (b) 2D square lattice. The
next-nearest-neighbor interaction in a square lattice is denoted
by $J$. }\label{fig:loop}
\end{figure}
In 2D square lattice, and also in  a lattice in any dimension
 where the minimum closed loop has a number of sides greater than three, the first non-zero
 contribution above the linear order in $V$ occurs at order $V^3$.
Interestingly, this can already be seen at the level of second
order cumulant expansion of the interactions. This is illustrated
by Fig. (1b). Note that in the square lattice, the first term in
Eq.~(\ref{secondorder}), upon the usual gradient expansion,
generates a next-nearest-neighbor interaction $J$. In a consistent
RG-treatment, $J$ (and in fact all such longer-ranged Josephson
interactions) should be included in the starting action. As has
been shown before, they are all irrelevant in the floating phase
as long as $V$ is irrelevant. However, as soon as $V$ becomes
relevant at $\alpha_c$, their recursion relations pick up source
terms of order $V^2$. They  in turn generate higher order
corrections to the recursion relations for $V$.

It is clear that the lowest order correction comes from the
next-nearest-neighbor interaction $J$. Including this term in the
starting action in Eq.~(\ref{action1}), and with the help of Fig.
(1b) and Eq.~(\ref{Jsirg}), we find for the recursion relation for
$J$,
\begin{equation}
\frac{dJ}{dl}=(1-\frac{1.26}{z_0\alpha})J + (\frac{0.19
\tau_0^{'}}{z_0\alpha})V^2, \label{Jrg}
\end{equation}
where $\tau_0^{'}$ is defined via $\int_{-\tau}^{\beta}
d(\tau^{'}-\tau)[\exp(-G^{f}_{12,23}(\tau-\tau^{'}))-1]=\tau_0^{'}[\exp(-G^{f}_{12,23}(0))-1]$.
By an exactly analogous calculation, we get for the recursion
relation for $V$,
\begin{equation}
\frac{dV}{dl}=(1-\frac{1}{z_0\alpha})V +
(\frac{0.63\tau_0^{''}}{z_0\alpha})VJ, \label{Vrg2}
\end{equation}
where $\tau_0^{''}$ is defined by a third integral,
$\int_{-\tau}^{\beta}
d(\tau^{'}-\tau)[\exp(-G^{f}_{13,34}(\tau-\tau^{'}))-1]=\tau_0^{''}[\exp(-G^{f}_{13,34}(0))-1]$.

Solving for $\frac{dJ}{dl}=0$ at $\alpha=\alpha_c$, we find $J\sim
V^2$, and putting it back in Eq.~(\ref{Vrg2}) we get,
\begin{equation}
\frac{dV}{dl}=(1-\frac{1}{z_0\alpha})V + C_2V^3,\label{Vrg3}
\end{equation}
where $C_2$ is a positive constant given in terms of the time
cut-offs, $C_2= (\frac{0.46\tau_0^{'}\tau_0^{''}}{z_0^2
\alpha\alpha_c})$.

There is a possibility of an order $V^3$ term appearing in the
recursion relation for $V$ at third order expansion of the
interactions  for the square lattice.
 This contribution may come from a term of the form

 \begin{eqnarray}
 &&\frac{V^3}{24}\int_0^\beta d\tau_1\int_0^\beta d\tau_2\int_0^\beta
 d\tau_3\sum_{\langle i, j \rangle}\sum_{\langle k, l \rangle}\sum_{\langle m, n \rangle}\cos[\Delta\theta^s_{ij}(\tau_1)
 \nonumber\\&+&\Delta\theta^s_{kl}(\tau_2)+\Delta\theta^s_{mn}(\tau_3)]
 e^{-\frac{3\ln
 b}{z_0\alpha}}[\exp(G^{f}_{ij,kl}(\tau_1-\tau_2)\nonumber\\&+&
 G^{f}_{kl,mn}(\tau_2-\tau_3)+G^{f}_{mn,ij}(\tau_3-\tau_1))-1]\nonumber,
 \label{thirdorder}
 \end{eqnarray}
which appears in the third order cumulant expansion of
Eq.~(\ref{partition}).
 We have checked explicitly, as shown in the appendix~\ref{app:RG}, that this does not
 happen.
 In fact, the contribution to Eq.~(\ref{Vrg3}) from third
 order cumulant expansion is precisely zero. Thus there is no cancellation of
 the order $V^3$ term we have found above. Note that for any
 lattice, where the minimum closed loop has number of sides
 greater than three, the order $V^3$ term appears via the
 generation of the
 next-nearest-neighbor interaction in the RG. Thus 2D honeycomb
 lattice, for example, will also have a recursion relation for $V$
 similar to Eq.~(\ref{Vrg3}).

 Combined with the flow equation for $\alpha$,
 $\frac{d\alpha}{dl}=0$, which is true in our
 perturbative treatment close to $\alpha_c$ for all lattice types,
 we have sketched in Fig.~\ref{fig:flow} the perturbative RG flows for 1D, 2D triangular
 lattice and 2D square lattice.

 \section{Experimental Consequences}
 \label{sec:experiments}
Certain aspects of our theory can be seen  in a recent
experiment \cite{Takahide} (see also references therein), although
some caution must be exercised. In this paper, a phase transition
is shown to exist at $\alpha_{c}=1/ z_{0}$, for small values of
$V/E_{0}$, where $z_{0}=2$ for the square lattice fabricated in
the experiment. This is consistent with the predictions first
given in Ref. \onlinecite{CIKZ} and later in Ref. \onlinecite{Fisher} from a
single junction approach. Moreover, the experimental  phase
diagram, especially the absence of a dissipation-driven transition
above a certain value of $V/E_0$, bears a remarkable resemblance
with what can be trivially inferred from our Fig. \ref{fig:flow}.
There are two issues that deserve further attention: (1) the
quantum disordered state of the resistively shunted Josephson
junction array is metallic, not an insulator, as inferred in this
experiment. This is simply because the metallic shunts will short
the current when the superconducting phase difference between the
junctions become incoherent, unless disorder effects in two
dimensions turn this metallic behavior to an insulating behavior.
However, the experimental system  is claimed to be free of
disorder. This metallic behavior was not tracked down in this
experiment, but only the insulating behavior of the unshunted
array; further experiments  can clearly be revealing; (2) the
correct model of the capacitances must also include longer ranged
contributions, as they are obviously present in the experiment.

 Given the novelty of the floating phase, further experiments
will be a worthwhile effort.
For values of $R>R_Q=\frac{h}{4e^2}$,
 we expect the system to be in the floating phase, where the
  grains are only short-range-correlated in space.
 We expect the current-voltage characteristic in
this case to be a non-universal power law controlled by
dissipation $\alpha$, similar to the results obtained previously
\cite{Kane} in the context of isolated junctions.  For $R$ smaller
than the quantum of resistance, the system will be in the
superconducting state, and the temperature-dependent non-universal
power law will be similar to the classical XY model due to vortex
unbinding in the presence of an imposed current. \cite{Kadin}
Recall that $V$, along with all other longer range Josephson
couplings, \cite{Sumanta1} flow to larger values in this regime.
An experimentally testable prediction for a non-universal exponent
$\nu$ controlling the critical scaling of a characteristic
temperature $T_{ch}$, $T_{ch} \propto(\delta V)^\nu$ with $\delta
V \equiv |V - V_c|$ and $V_c$ the critical value of $V$ on the
($V-\alpha$) plane, was made in our earlier work. \cite{Sumanta1}
 If verified, this will be a clear cut signature of the phase
 transition studied here. By
varying the strengths of the Josephson coupling and the shunt
resistors, the critical locus and hence the phase boundary on the
($V-\alpha$) plane can also be directly examined.

The ideas presented here assume significance also in a broader
context. \cite{Varma, Si} In a bid to understand some puzzling
experimental observations in the context of heavy fermion
materials, \cite{Heavyfermions} and the cuprate superconductors,
\cite{Cuprates} QPT's are sought where extended systems may go
through global phase transitions which are ultimately driven by
inherent local ones. The transition we have studied in the
dissipative Josephson junction array provides just such an
example, albeit only in spirit, since we have not treated the
fermions here. It will be interesting to extend the ideas and
results found here to fermionic problems to make contact with a
wide range of systems of experimental interest.

\section{Summary and Conclusion}
\label{sec:conclusion} In this paper, we have  clarified in detail
the nature of the correlation functions within, and the transition
into, the floating phase of a dissipative Josephson junction
array.
%where the nearest neighbor and all other longer range
%Josephson interactions are irrelevant. \cite{Sumanta1}
%Although
%the only other irrelevant variable $C$ does not affect the
%recursion relations, it is dangerously irrelevant, and gives rise
%to interesting correlations in the floating phase.
We have shown that the long-time correlations in the floating
phase are finite locally for each superconducting grain, but the
spatially long-range correlations are identically zero at $T=0$.
At any nonzero temperature, the spatial correlation functions are
exponentially decaying with a finite correlation length,
indicating locally dynamic grains which are only short-range
correlated in space. These correlations help identify the
dissipation-driven metal to superconductor transition in the
extended system of the array with the `quantum to classical'
dynamic transition existing at the level of a single junction;
\cite{Chakravarty1, Bray} the local transition drives the global
one.

Further, we have shown that the transition in the extended system
 is controlled by a locus of critical points in the $\alpha-V$
plane with continuously varying critical exponents. To derive our
results, we computed the perturbative recursion relations for the
nearest neighbor coupling $V$ up to third order in $V$. Since $V$
is the assumed small parameter, the calculation is valid only
close to the critical value of $\alpha=\alpha_c$ needed for
ordering with infinitesimal Josephson coupling. We have found no
correction to the linear-order recursion relation, Eq.~\ref{Vrg},
for $D=1$. In this case, then, the fixed line close to $\alpha_c$
is parallel to the $V$-axis. This is a result similar to that
obtained for the corresponding zero-dimensional problem treated in
Ref.\onlinecite{Zwerger}. Here we have extended their result to
$D=1$.
%Note that with our capacitance term $C$ present, the
%one-dimensional problem does not map on to the zero-dimensional
%one in any obvious way.

For any $D>1$, however, there are higher order corrections to
Eq.~\ref{Vrg}. Coupled with the recursion relation for $\alpha$,
$\frac{d\alpha}{dl}=0$, the locus of critical points bends toward
the $V$-axis determining a phase boundary. This locus is a
straight line in any $D>1$ as long as the number of sides in a
minimum closed loop is three, it's a parabola if that number is
greater than three. The functional form of the locus, which
determines the phase boundary between the metal and the
superconductor, is entirely determined by the {\it local} topology
of the lattice, and not on the global dimension. Consequently, the
locus of critical points exists in any dimension; there is no
upper-critical-dimension for the problem. All of these results are
consistent with the global transition being driven by the inherent
local one.

 The correlations in the floating phase and the phase boundary
  studied here can be experimentally
 probed by measuring the current-voltage characteristics with the
 resistance of the shunt resistors continuously tuned.
 In a broader context, taken together with
our earlier result \cite{Sumanta1} that all longer-range Josephson
interactions induced by the lattice among a set of periodically
spaced local systems, which each goes through individual local
transitions with dissipation as the tuning parameter, are
irrelevant in the RG sense, our results here assume significance
in the context of the search for local quantum criticality in
extended systems. \cite{Varma}

%We have also discussed in detail the  correlation functions in the
%disordered floating phase, where $V$ and all other longer range
%interactions are irrelevant.  Although the only other irrelevant
%variable $C$ does not affect the recursion relations, it is
%dangerously irrelevant, and gives rise to interesting correlations
%in the floating phase. We have shown that the long-time
%correlations are finite locally for each superconducting grain,
%but the spatially long-range correlations are identically zero at
%$T=0$. At any nonzero temperature, the spatial correlation
%functions are exponentially decaying with a finte correlation
%length indicating locally dynamic grains which are only
%short-range correlated in space.

 %To summarize, for the dissipative Josephson-junction
 %array, a locus of critical fixed points
 %with continuously varying exponents exists in any dimension, it
 %bends towards the $\alpha=0$ axis determining a phase-boundary
 %between the metal and the superconductor for any $D>1$, and its
 %functional form only depends on the local topology of the lattice,
 %but not on the global dimension.

 \section{Acknowledgements}
 This work was supported by the NSF under
grants:  DMR-0411931, DMR-0132555 and DMR-0132726. We thank S.
Kivelson, B. Spivak, and T. Kirkpatrick for discussions, and  the
Aspen Center for Physics for their hospitality while a portion of
this work was being done. S. T. also thanks the Department of
Physics and the Institute of Theoretical Science at the University
of Oregon for their hospitality when a part of this work was being
completed.

\appendix
\section{Spatial correlation functions in the floating phase}
\label{app:correlation} To compute the spatial correlation
functions of the phase variables at a temperature
$\frac{1}{\beta}$, we need to evaluate the following integral in
frequency and wavenumber space:
\begin{equation}
\langle\theta_i(\tau)\theta_{i+r}(\tau)\rangle=\sum_{\frac{1}{\beta}\leq|\omega_n|}\frac{1}{|\omega_n|}
\int_0^{\Lambda}\frac{d^dk}{(2\pi)^d}\frac{\exp(i{\bf k}.{\bf
r})}{\alpha' f({\bf k})+C|\omega_n|} \label{integral}
\end{equation}
where $\alpha'=\alpha/2\pi$ for notational convenience. Noticing
that only $k \sim 1/r$ modes will contribute significantly to the
integral, and for large values of $r$, we expand the function
$f({\bf k})\sim z_0k^2$ in the denominator. The $k$-space integral
is then simply the Fourier-transform of the Yukawa potential at
wavenumber space.
\subsection{Correlation function in D=3}
In $D=3$, doing the $k$-space integral, which simply gives the
three dimensional Yukawa potential, and after a change of
variable, we get
\begin{equation}
\langle\theta_i(\tau)\theta_{i+r}(\tau)\rangle=\frac{1}{2\pi^2\alpha
'z_0}\frac{1}{r}\int_{r/\xi}^{\infty}\frac{du}{u}\exp(-u),
\label{3D1}
\end{equation}
where $\xi^{-1}\equiv\xi(T)^{-1}=\surd(\frac{CT}{\alpha' z_0})$.
The $u$-integral produces the exponential-integral function,
\cite{Table} and the exact correlation function is given by,
\begin{equation}
\langle\theta_i(\tau)\theta_{i+r}(\tau)\rangle=-\frac{1}{2\pi^2\alpha
'z_0}\frac{1}{r}Ei(-r/\xi). \label{3D2}
\end{equation}
Taking the asymptotic values of $Ei(x)$ for large and small $|x|$,
we end up with Eq.~\ref{3Dscaling}. With the help of the result
$\langle\theta_i(\tau)^2\rangle=\frac{\Lambda}{2\pi^2\alpha'z_0}\ln(\beta)$,
we get Eq.~\ref{3Dopcorr}.
\subsection{Correlation function in D=2}
In $D=2$, the angular integration in $k$-space produces $J_0(kr)$,
where $J_0(z)$ is the Bessel function of the first kind of  order
zero. \cite{Table} The remaining part of the $k$-space integral
produces the function $2\pi K_0((\frac{C\omega}{\alpha '
z_0})^{\frac{1}{2}}r)$, where $K_0(z)$ is simply related to the
Hankel function of the first kind $H_0^{1}(z)$. After a change of
variable, we end up with the integral,
\begin{equation}
\langle\theta_i(\tau)\theta_{i+r}(\tau)\rangle=\frac{1}{\pi^2
\alpha ' z_0}\int_{r/\xi}^{\infty}\frac{du}{u}K_0(u). \label{2D1}
\end{equation}
For $r>>\xi$, we use the asymptotic expansion of the function
$K_0(u)$ for large $u$, and after doing the $u$-integral we get
the result
\begin{eqnarray}
\langle\theta_i(\tau)\theta_{i+r}(\tau)\rangle_{r>>\xi}&=&\frac{1}{\pi^2
\alpha'
z_0}\surd(\frac{\pi}{2})(\frac{\xi}{r})^{3/4}\exp(-r/2\xi)\nonumber\\&\times&W_{-3/4,-1/4}(r/\xi),
\label{2D2}
\end{eqnarray}
where $W_{\lambda,\mu}(z)$ is the Whittaker function. \cite{Table}
Asymptotically expanding the Whittaker function for $r>>\xi$, we
end up with the first line of Eq.~\ref{2Dscaling}. In the limit
$r<<\xi$, we notice that the integral in Eq.~\ref{2D1} is
dominated by small $u$. Expanding $K_0(u)$ for small u, and doing
the $u$-integral we get the remaining part of Eq.~\ref{2Dscaling}.
With the additional result in $D=2$,
$\langle\theta_i(\tau)^2\rangle=\frac{1}{8\pi^2\alpha 'z_0}(\ln
\frac{\beta\alpha 'z_0}{C})^2$, we finally get Eq.~\ref{2Dopcorr}.
\subsection{Correlation function in D=1}
In $D=1$, after doing the $k$-space integral and a change of
variable, we get,
\begin{eqnarray}
\langle\theta_i(\tau)\theta_{i+r}(\tau)\rangle&=&\frac{r}{\pi
\alpha '
z_0}\int_{r/\xi}^{\infty}\frac{du}{u^2}\exp(-u)\nonumber\\
&=&\frac{r}{\pi \alpha '
z_0}[Ei(\frac{-r}{\xi})+\frac{\xi}{r}e^{-\frac{r}{\xi}}].
\label{1D1}
\end{eqnarray}
 Asymptotically expanding  the exponential-integral function for
large $r/\xi$, the first term in the expansion cancels the second
term within the square bracket of Eq.~\ref{1D1}. The first
subleading term in the expansion produces the first line in
Eq.~\ref{1Dscaling}. Expansion of $Ei(r/\xi)$ for small values of
the argument produces the remaining part. With the additional
result that in $D=1$, $<\theta_i(\tau)^2>=\frac{1}{\pi\surd(\alpha
'z_0C)}\beta^{1/2}$, we end up with Eq.~\ref{1Dopcorr}.

Finally, by doing the change of variables $\beta\omega=\Omega$ and
${\bf k}r={\bf Q}$ where $r$ is the magnitude of the spatial
coordinate ${\bf r}$, and taking $1/(\alpha ' z_0)$ outside the
integral in Eq.~\ref{integral}, we get the scaling relation
Eq.~\ref{scaling}.

\section { RG from the third order cumulant expansion in $V$}
\label{app:RG} Here we show that the term of order $V^3$ in the
recursion relation for $V$, found from the cumulant expansion of
the partition function of Eq.~\ref{partition} at third order,
identically vanishes for any lattices in arbitrary dimensions.
Using the formula for third order cumulant expansion,
\begin{equation}
 \langle\exp[f]\rangle=\exp[\langle f \rangle + \frac{1}{2}(\langle f^2\rangle -\langle f
\rangle^2)(1-\langle f \rangle)+\frac{\langle f^3
\rangle}{6}-\frac{\langle f \rangle ^3}{6},
\end{equation}
We get, for $f=V\int_{0}^{\beta}d\tau\sum_{\langle i,j
\rangle}\cos\Delta\theta_{ij}(\tau)$,
\begin{widetext}
\begin{eqnarray}
\exp[f]&=&\exp\Big[Ve^{-\ln
b/z_0\alpha}\int_0^{\beta}d\tau\sum_{\langle
i,j\rangle}\cos\Delta\theta_{ij}^s(\tau)+\Big(
\frac{V^2}{4}e^{-2\ln b/z_0\alpha }\sum_{\langle i,j
\rangle}\sum_{\langle k,l
\rangle}\int_0^{\beta}d\tau_1\int_{0}^{\beta}d\tau_2\Big[\cos(\Delta\theta_{ij}^s
(\tau_1)+\Delta\theta_{kl}^s(\tau_2))\nonumber\\&\times&(e^{-G^{f}_{ijkl}(\tau_1-\tau_2)}-1)+
\cos(\Delta\theta_{ij}^s
(\tau_1)-\Delta\theta_{kl}^s(\tau_2))\times(e^{G^{f}_{ijkl}(\tau_1-\tau_2)}-1)\Big]\Big)\Big(
1-Ve^{-\ln b/z_0\alpha}\int_0^{\beta}d\tau_3\sum_{\langle
m,n\rangle}\cos\Delta\theta_{mn}^s(\tau_3)\Big)\nonumber\\&+&\frac{\langle
f^3\rangle_{0f}}{6} - \frac{\langle f \rangle_{0f}^3}{6}\Big]
\label{expansion}
\end{eqnarray}
\end{widetext}

After evaluating the correlation functions of the fast modes,
$G^{f}_{ijkl}(\tau_1\-\tau_2)$'s, doing gradient expansions in
$(\tau_2-\tau_1)$ and $(\tau_3-\tau_1)$ (it can be seen by
straightforwardly writing out ($\frac{\langle f^3\rangle_{0f}}{6}
- \frac{\langle f \rangle_{0f}^3}{6}$) in the above expression
that, all terms where $\tau_1, \tau_2$ and $\tau_3$ are not
constrained to be close to each other cancel among themselves),
and finally rescaling $\tau_1$, a general term that may contribute
to renormalize $V$ at third order has the form (with $\ln b=l$),
\begin{eqnarray}
&&V^3e^{l}e^{-3l/z_0\alpha}\int_0^{\beta}d\tau_1\sum_{\langle i, j
\rangle}\sum_{\langle k, l \rangle}\sum_{\langle m, n \rangle}
\cos\Big(\Delta\theta_{ij}(\tau_1)\nonumber\\&+&\Delta\theta_{kl}(\tau_1)+\Delta\theta_{mn}(\tau_1)\Big)
\times\int_{-\tau_1}^{\beta}d(\tau_2-\tau_1)\nonumber\\&&\int_{-\tau_1}^{\beta}d(\tau_3-\tau_1)
\Big(\exp[-G^f_{ijkl}(\tau_2-\tau_1)-G^f_{klmn}(\tau_3-\tau_2)\nonumber\\&-&G^f_{mnij}(\tau_3-\tau_1)]-1\Big)
\end{eqnarray}
 Note that the
$G^{f}_{ijkl}(\tau_1\-\tau_2)$'s  are evaluated within a thin
shell $\omega_c e^{-l}<|\omega|<\omega_c$ around the high
frequency cut-off. Using the property that they go to zero as
$l\rightarrow 0$, we have to expand their exponentials appearing
in the above expansion in linear order in the infinitesimal
parameter $l$ to get their contributions to the recursion relation
of $V$. Hence, we do this expansion in $l$ right away in all of
the terms appearing in Eq.~\ref{expansion}. We also use the
trigonometric identity
$\cos(A)\cos(B)=\frac{1}{2}[\cos(A+B)+\cos(A-B)]$ to simplify the
second term of the argument of the exponential in
Eq.~\ref{expansion}. Then writing out ($\frac{\langle
f^3\rangle_{0f}}{6} - \frac{\langle f \rangle_{0f}^3}{6}$) using
three variables $\epsilon_1, \epsilon_2, \epsilon_3$, which take
the values $\pm 1$, performing the averages with respect to the
fast modes (note that no higher order correlation functions arise,
only the familiar $G^{f}_{ijkl}(\tau_1-\tau_2)$ with various
combinations of the indices suffice), and using all eight
permutations of the values of the $\epsilon$-variables, we get
sixteen different terms at order $V^3$. Denoting
$\cos\Big(\Delta\theta_{ij}(\tau_1)+\Delta\theta_{kl}(\tau_2)+\Delta\theta_{mn}(\tau_3)\Big)$
by $(ij + kl + mn)$ and so on, and omitting the summation and
integration signs along with a multiplicative factor $\exp(-3\ln
b/z_0\alpha)$, they are,
\begin{widetext}
\begin{eqnarray}
&&\frac{V^3}{8}\Big[
(ij+kl+mn)\Big(G^f_{ijkl}(\tau_1-\tau_2)-\frac{1}{3}
G^f_{ijkl}(\tau_1-\tau_2)-\frac{1}{3}G^f_{klmn}(\tau_2-\tau_3)-\frac{1}{3}G^f_{mnij}(\tau_3-\tau_1)\Big)\nonumber\\&+&
\Big(\frac{2}{3}(ij+kl-mn)G^{f}_{ijkl}(\tau_1-\tau_2)-\frac{1}{3}(kl+mn-ij)G^f_{klmn}(\tau_2-\tau_3)
-\frac{1}{3}(mn+ij-kl)G^f_{mnij}(\tau_3-\tau_1)\Big)\nonumber\\
&-&
\Big(\frac{4}{3}(ij-kl+mn)G^{f}_{ijkl}(\tau_1-\tau_2)-\frac{1}{3}(kl+mn-ij)
(G^f_{ijkl}(\tau_1-\tau_2)+G^f_{mnij}(\tau_3-\tau_1))
-\frac{1}{3}(ij+kl-mn)(G^f_{klmn}(\tau_2-\tau_3)\nonumber\\&+&G^f_{mnij}(\tau_3-\tau_1))\Big)\Big].
\label{expansion1}
\end{eqnarray}

\end{widetext}

 Keeping track of the signs, using the pairwise symmetries among
the indices ($\langle i, j \rangle, \langle k, l \rangle, \langle
m, n \rangle$), and using the property that
$G^f_{ijkl}(\tau_1-\tau_2)$ is even in space and time, it is
straightforward to see that all of the terms in
Eq.~\ref{expansion1} cancel among themselves. Hence, the third
order cumulant expansion of Eq.~\ref{partition} does not
renormalize $V$. Note that, to get this result, we did not have to
assume any partcular lattice-type; it's generally true for all
lattices in any dimension. Among other things, we have thus
confirmed the conjecture made in Ref.~\onlinecite{Zwerger} about
vanishing of the third order term in the corresponding
zero-dimensional problem. Moreover, we have established that the
same is true in any higher dimension. This calculation does not
imply, however, that there
    is no order $V^3$ correction to Eq.~\ref{Vrg}. As we have
    seen in Eq.~\ref{Vrg3}, the order $V^3$ term arises already at
    the second order cumulant level, by generating the
    diagonal Josephson interaction $J$ in a square lattice. In
    $D=1$, however, there is no higher order correction to
    Eq.~\ref{Vrg}.

\end{document}